\documentclass[twocolumn, amsmath,amssymb,aps, prl,
showkeys,showpacs, floatfix, superscriptaddress] {revtex4-2}

\usepackage{times}
\usepackage{graphicx}
\usepackage[usenames,dvipsnames]{color}
\usepackage{bm}
\usepackage{amsmath}
\usepackage{amssymb}

\usepackage[colorlinks=true,breaklinks=true,allcolors=blue]{hyperref}

\begin{document}

\title{Universal Euler-Cartan Circuits for Quantum Field Theories} 
\author{Ananda Roy}
\email{ananda.roy@physics.rutgers.edu}
\affiliation{Department of Physics and Astronomy, Rutgers University, Piscataway, NJ 08854-8019 USA}
\author{Robert M. Konik}
\affiliation{Division of Condensed Matter Physics and Material Science, Brookhaven National Laboratory, Upton, NY 11973-5000, USA}
\author{David Rogerson}
\affiliation{Department of Physics and Astronomy, Rutgers University, Piscataway, NJ 08854-8019 USA}

\begin{abstract}
Quantum computers can efficiently solve problems which are widely believed to lie beyond the reach of classical computers. In the near-term, hybrid quantum-classical algorithms, which efficiently embed quantum hardware in classical frameworks, are crucial in bridging the vast divide in the performance of the purely-quantum algorithms and their classical counterparts. Here, a hybrid quantum-classical algorithm is presented for the computation of non-perturbative characteristics of quantum field theories. The presented algorithm relies on a universal parametrized quantum circuit ansatz based on Euler and Cartan's decompositions of single and two-qubit operators. It is benchmarked by computing the energy spectra of lattice realizations of quantum field theories with both short and long range interactions. Low depth circuits are provided for false vacua as well as highly excited states corresponding to mesonic and baryonic excitations occurring in the analyzed models. The described algorithm opens a hitherto-unexplored avenue for the investigation of mass-ratios, scattering amplitudes and false-vacuum decays in quantum field theories.
\end{abstract}

\maketitle

\paragraph*{Introduction.}
Quantum field theory~(QFT), which reconciles quantum mechanics and the special theory of relativity, is fundamental to the description of a vast array of physical phenomena. The latter range from confinement and asymptotic freedom in theories of elementary particles to high-temperature superconductivity and topological order in condensed matter systems. The investigation of non-perturbative characteristics of generic QFTs remains an outstanding challenge with ab-initio computations likely to stay beyond the reach of the most powerful classical computers. Quantum computation~\cite{Feynman_1982} provides an alternate paradigm for the analysis of QFTs with quantum algorithms boasting exponential speedup over classical ones for the computation of scattering amplitudes in massive scalar~\cite{Jordan2011, Jordan2012} and fermionic~\cite{Jordan2014} theories. 

Despite their potential, it is a daunting challenge to implement quantum algorithms on state-of-the-art quantum simulators and solve QFT problems of widespread interest. This is because the current quantum simulators contain only a few hundred qubits with modest error rates. In the current noisy quantum era, hybrid quantum-classical algorithms, which combine the power of a quantum processor with that of a classical one, are crucial in bridging the enormous divide between purely-quantum algorithms and their classical counterparts. Notable examples include the variational quantum eigensolver~\cite{Peruzzo:2013bzg, Tilly2022} and its modifications~\cite{Grimsley2019, Tang2021} as well as the quantum approximate optimization algorithm~\cite{Farhi2014, Lloyd2018}, which have been successful in analyzing problems in quantum chemistry~\cite{Kandala2017} and combinatorial optimization~\cite{Harrigan2021}. 

There are two main elements of a hybrid algorithm that govern its efficacy: i) a parametrized quantum circuit ansatz built with a predetermined pool of unitary operators, ii) an efficient optimization procedure for the parameters specifying the unitary operators. While the second element is typically based on variations of gradient descent methods~\cite{Fletcher2013, Kingma2017,Stokes2020},  there is no natural prescription for the first. The circuit ansatz either relies on the Hamiltonian whose eigenstate is being sought alongside a heuristically-chosen mixer Hamiltonian~\cite{Farhi2014, Hadfield2019, Wang2020} or variations of the coupled cluster theory relevant mostly for chemistry problems~\cite{Peruzzo:2013bzg}. Neither of the aforementioned are suitable for generic QFT problems whose Hamiltonians, when mapped to registers of qubits in a quantum computer, could involve long-range couplings~\cite{Banuls:2013jaa, Farrell2024} or might not even be Hermitian~\cite{Ikhlef2012, Bazhanov:2020dlm}. Furthermore, many interesting properties of QFTs such as mass-ratios and scattering amplitudes remain encoded in highly-excited states~\cite{Luscher1990, Luscher1991, Hardy:2024ric}, which have largely eluded existing hybrid algorithms. Last but not least, restricting the circuit ansatz to a given pool of unitary operators that does not encompass the entire manifold of possible unitary operators necessarily implies that the result so obtained may not be the globally optimum solution which could potentially be found by considering a larger ansatz. 

\begin{figure*}
\centering
\includegraphics[width=\textwidth]{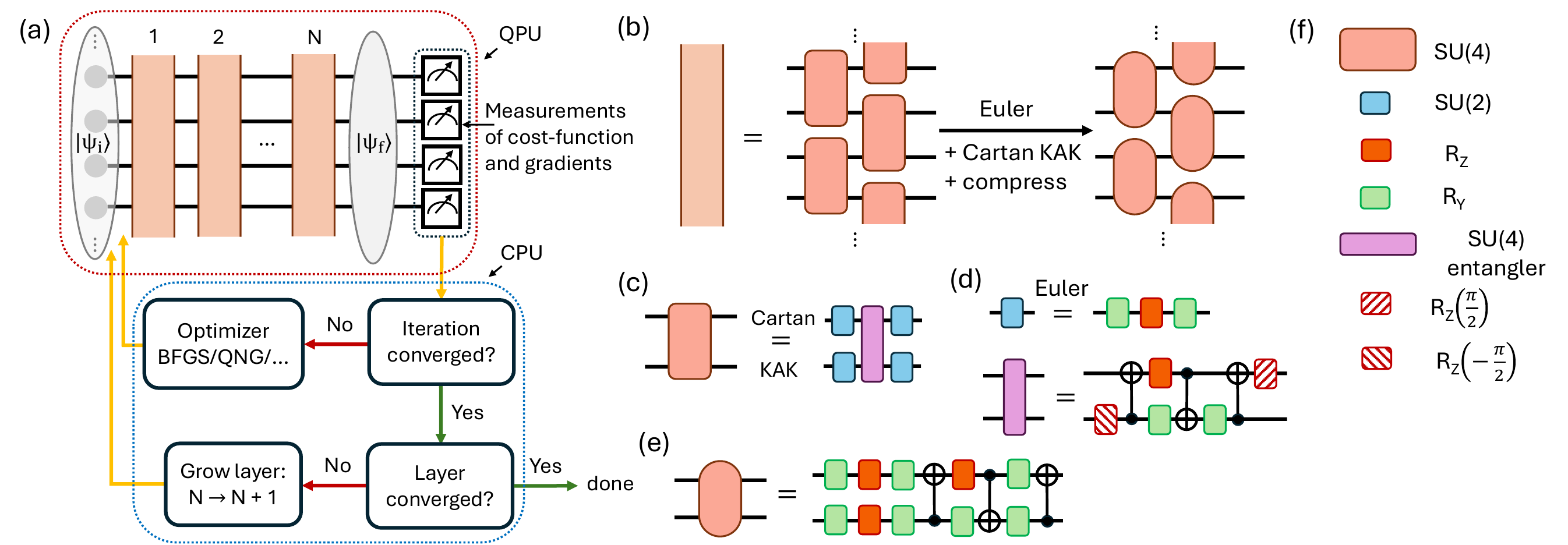}
\caption{\label{fig:schematic} (a) Schematic of the proposed algorithm initialized with a state~$|{\rm \psi_i}\rangle$. The quantum processor~(maroon dotted line) is involved in the application of~$N$ layers of parametrized unitary operators yielding the state~$|{\rm \psi_f}\rangle$, followed by measurements of a cost-function and relevant gradients. A classical processor~(blue dotted line) verifies the iteration convergence criterion. This checks if the results for the cost-function are converged when compared to that obtained in the previous iteration step for a given $N$. If the iteration convergence criterion is not met, the parameters of the~$N$ unitary operators are updated using an optimization routine. Some typical examples are BFGS and the quantum natural gradient~(QNG). The updated set of parameters are used in the next iteration and this process is repeated until the desired iteration convergence criterion is reached. Subsequently, a second layer convergence criterion is checked by comparing the result to that obtained with~$N-1$ layers. If the convergence criterion is satisfied, the entire process terminates, else the number of layers is grown by 1 and the previous steps are repeated. (b) Decomposition of each layer unitary operator in terms of  general two-qubit unitary operators~(orange rectangles) and those after Euler and Cartan's KAK decompositions and compression~(orange ellipses). (c) Cartan's KAK decomposition of a general SU(4) two-qubit unitary operator in terms of four single-qubit SU(2) operators and the SU(4) entangler. (d) Euler's decomposition of a general SU(2) operator in terms of rotations by three angles about two non-parallel axes~(chosen here to be~$\hat{y}$ and~$\hat{z}$). The SU(4) entangler is decomposed into three controlled-NOT gates and five single-qubit rotations. (e) The two-qubit operator after Euler's and Cartan's decompositions used for the optimization process. (f) Legends for the different symbols used. The symbol~$R_\alpha$ stands for the SU(2) operator~$e^{-i\theta\alpha}$, where~$\alpha$ is one of the three Pauli matrices and~$\theta$ is an unspecified angle determined by the optimization procedure. Two boxes of the same color perform different amounts of rotations about the same axis.}
\end{figure*}

In light of the above, it is natural to ask: is there a quantum-classical algorithm based on a universal parameterized quantum circuit ansatz that is tractable with established optimization strategies and applicable to a broad set of QFT problems? The goal of this work is to answer this question. The presented solution computes the optimal circuit over all possible two-qubit unitary rotations required to realize an arbitrary $L$-qubit unitary rotation~\cite{Nielsen_Chuang_2000}. The unitary operators are chosen to couple neighboring qubits, which is imposed for potential implementation on majority of quantum computing architectures and could be relaxed in a more general setting. The model under investigation affects the parameters of the circuit, but not the ansatz itself. Importantly, the proposed ansatz explores the largest possible manifold of allowed unitary operators and is closer to the globally optimal solution than what could be obtained by considering a more restrictive ansatz. The potential of the algorithm is demonstrated by computing not only ground states of strongly-interacting QFTs, but also highly excited states that correspond to bound-states of topological excitations occurring in these theories.

\paragraph*{Algorithm.} The algorithm is implemented on a quantum processor coupled with a classical one~(Fig.~\ref{fig:schematic}). The former is involved in  applying the parametrized circuit and conducting measurements required for the optimization process. The classical processor receives the measurement outcomes, performs the optimization and feeds the outcome back into its quantum counterpart. 

\begin{figure*}
\centering
\includegraphics[width=0.85\textwidth]{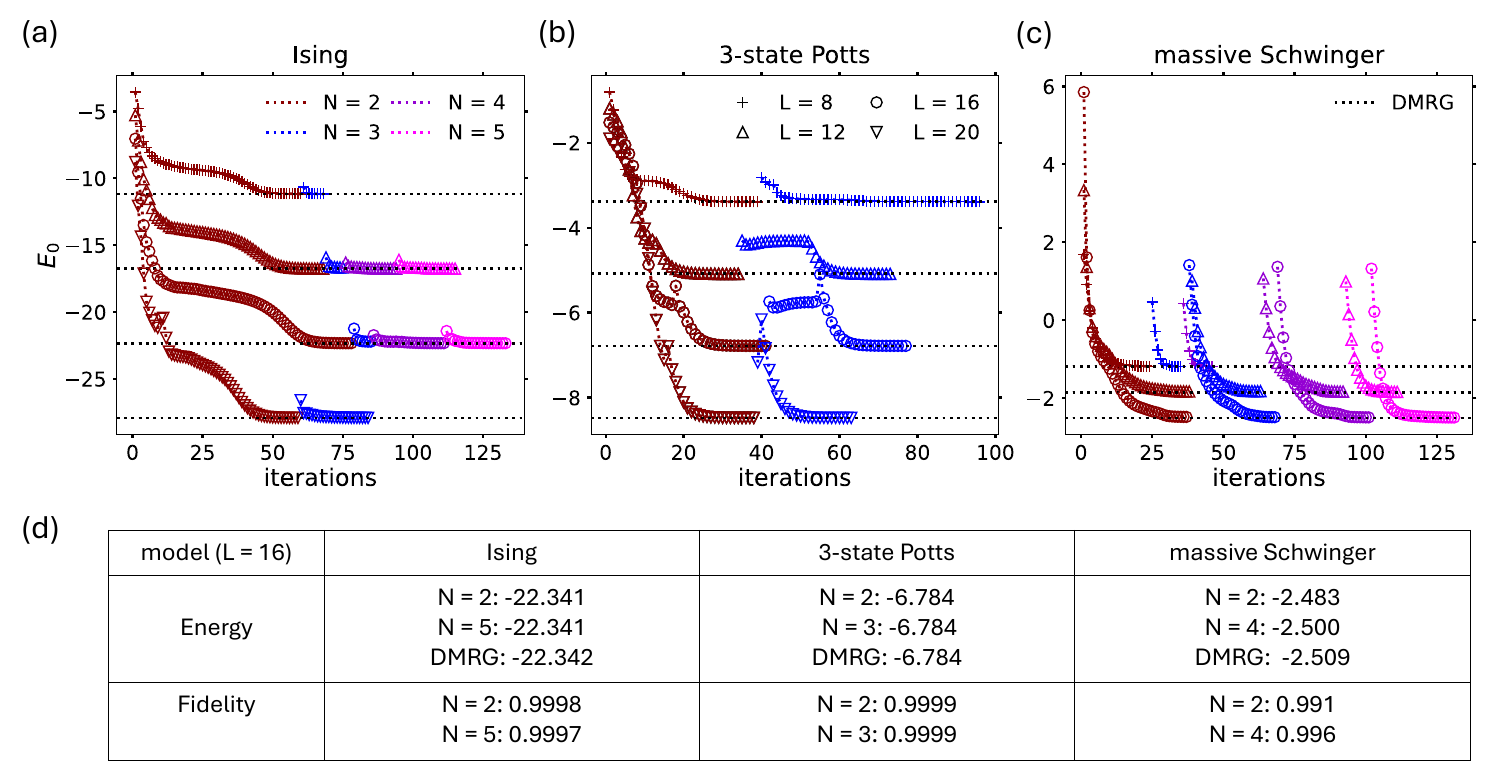}
\caption{\label{fig_2} (a) - (c) Results obtained for the ground state energies~($E_0$) of the Ising, 3-state Potts and the massive Schwinger models using the optimized Euler-Cartan circuits. A periodic chain of~$L$ qubits is subject to~$N$ layers for all three models. A translation-invariant Euler-Cartan circuit ansatz is optimized for panels (a, b), while the most general ansatz is considered in panel (c). The different values of $N~(L)$ are distinguished by the different colors~(markers). For the lowest value of~$N$, a uniform choice of $\theta_0 = 0.1$ was used as a starting guess for all of the angles being optimized. Once the iteration convergence criterion is met for a chosen value of $N$, the number of layers is increased by one:~$N\rightarrow N+1$. While optimizing for~$N+1$ layers, the previously obtained results for the~$N$ layers was used as a starting guess, augmented by an initial guess of $\theta_0/10$ for the last layer. This process was continued until the layer-convergence criterion was reached. Both convergence criteria were chosen to be $5\times10^{-4}$. The learning rates were chosen to be~$0.04$ for panels (a), (b) and 0.1 for panel (c). (d) Energies and fidelities to the corresponding target states for the smallest and largest values of~$N$ for~$L = 16$ are compared to the DMRG results. The smallest choice of~$N=2$ layers yielded results within $1\%$ error of the target energies, while the additional layers caused changes in the third and fourth significant digits. While the precision for the massive Schwinger model is lower due to the all-to-all couplings in the spin Hamiltonian~[Eq.~\eqref{eq:H_MS}] when compared to the other two models, increasing the number of layers leads to further improvement in the obtained results~(see also Ref.~\cite{Rogerson2024}). Note that the exact values of the changes in the energies with layers depend on the specific choices of the coupling constants and the distances between the initial and target ground states. }
\end{figure*}

\begin{figure*}
\centering
\includegraphics[width=0.8\textwidth]{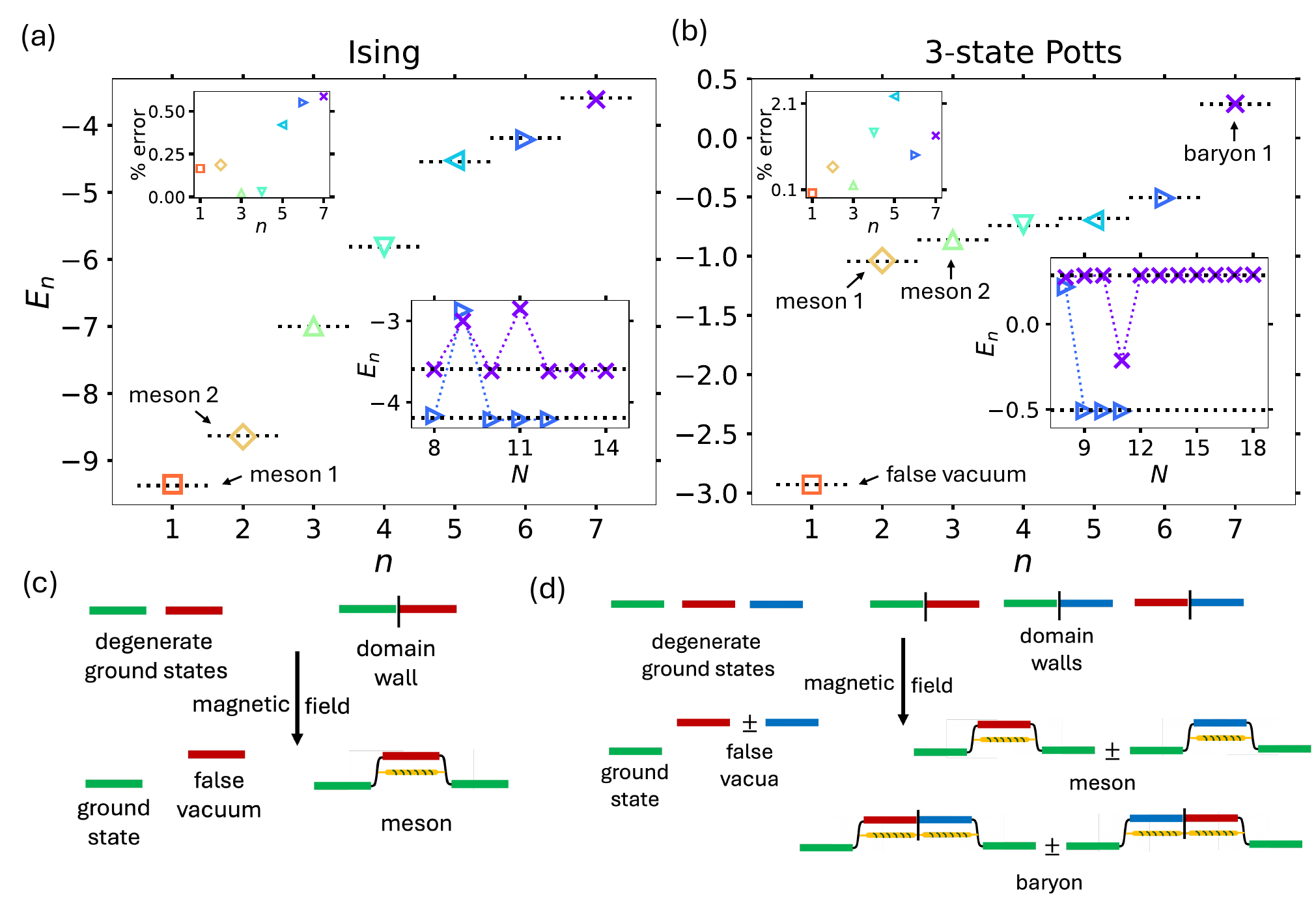}
\caption{\label{fig_3} (a) - (b) Results for the first seven excited states obtained using optimized Euler-Cartan circuits for the Ising [(a)] and Potts [(b)] models encoded in a periodic chain of~$L = 8$ qubits. The error percentages when compared to exact diagonalization results (black dotted lines) are shown in the top left insets. The variation of the obtained energies with number of layers is shown in the bottom right insets. The energies were obtained with a convergence criterion of $10^{-3}(10^{-4})$ for the Ising~(Potts) models with the same learning rate of 0.02. In both cases, a translation-invariant circuit ansatz was used. The Euler-Cartan circuit for the~$i^{\rm th}$ state was obtained by minimizing the Hamiltonian cost-function while orthogonalizing against the previously obtained states. Furthermore, in the Potts model, another term was added to the cost-function to obtain states  with~$+1$ eigenvalue for the operator~$C$~(see maintext). The seventh translation-invariant excited state in the Ising~(Potts) models corresponds to the $48^{\rm th}(42^{\rm nd})$ excited state in the spectrum. (c) - (d) Schematic of the confinement phenomena in the Ising and Potts models in the presence of a magnetic field. The latter breaks the two~(three)-fold degeneracy of the ground state leading to one true ground state and one~(two) false vacua in the Ising~(Potts) case. As a consequence, the domain-walls experience a confining force~(yellow lines with stripes), leading to the formation of mesonic and baryonic excitations. The non-monotonicity in the convergence for highly excited states can be improved by lowering the learning rates used in the circuit optimization process. }
\end{figure*}

The algorithm begins with the application of $N$~(see Figs.~\ref{fig_2}, \ref{fig_3} for representative choices of $N$) layers of unitary operators on~$L$ qubits of the quantum processor initialized in the state~$|\psi_{\rm i}\rangle$. Each of the~$N$ layers is built, in principle, out of two sublayers of two-qubit operators~[orange rectangles in Fig.~\ref{fig:schematic}(b)]. Each two-qubit operator is an element of the group SU(4). It can be written, using Cartan's KAK decomposition~\cite{Helgason1978, Tucci2005}, as a product of four single-qubit SU(2) rotations and an SU(4)-entangler~[Fig.~\ref{fig:schematic}(c)]. The latter when involving qubits~$i, i+1$ equals~$e^{-i\left(\alpha X_iX_{i+1} + \beta Y_iY_{i+1} + \gamma Z_iZ_{i+1}\right)}$ where~$X_i, Y_i$ and~$Z_i$ are the Pauli operators of the~$i^{\rm th}$ qubit. The SU(4) entangler can be realized by three controlled-NOT gates and five single-qubit rotations~\cite{Vatan2004, Vidal2004}~[Fig.~\ref{fig:schematic}(d)]. Finally, each of the four single-qubit rotations depicted in panel Fig.~\ref{fig:schematic}(c) can be further decomposed into three single qubit rotations by the well-known Euler angles~\cite{Nielsen_Chuang_2000}. Using the above, it can be shown that the general circuit ansatz is constructed with two-qubit operators~[orange ellipses in Fig.~\ref{fig:schematic}(b, e)] with ten undetermined angles. 

After the application of the~$N$ layers, measurements are performed to determine the relevant gradients and the cost-function~${\cal L}(\vec{\theta})$, where~$\vec{\theta}$ contains all parameters of the~$N$ layer quantum circuit. To obtain the ground state of a given Hamiltonian~$H$, the cost-function is chosen to be~${\cal L}(\vec{\theta}) = \langle \psi_{\rm f}(\vec{\theta})|H|\psi_{\rm f}(\vec{\theta})\rangle$. To obtain the~$n^{\rm th}$ excited state, the cost-function is augmented to~${\cal L}(\vec{\theta})\rightarrow {\cal L}(\vec{\theta}) + \sum_{m < n}\lambda_m|\langle \psi_m|\psi_{\rm f}\rangle|^2$, with~$\lambda_m>0, m = 0,\ldots, n-1$. This choice amounts to simultaneously minimizing the energy and the overlap to the~$n-1$ lower-energy states that have been already obtained.  

The measurement results are communicated to the classical processor which checks for the convergence of the cost-function for the given choice of~$N$~(iteration convergence in Fig.~\ref{fig:schematic}). If the condition is not satisfied, the parameters of the circuit are updated using an optimization routine. In this work, the quantum natural gradient optimizer~\cite{Stokes2020} is used due to its superior performance in obtaining eigenstates of many-body Hamiltonians~\cite{Wierichs2020, Roy2023efficient}. In contrast to ordinary gradient descent, this  ensures that optimization follows the steepest descent taking into account the geometry of the space of wavefunctions~\cite{Stokes2020, Zanardi2007} and can viewed as the quantum analog of the classical information geometric approach~\cite{Amari1998}. The parameters at the~$(t+1)^{\rm th}$ iteration are related to the~$t^{\rm th}$ one by
\begin{equation}
\vec{\theta}_{t + 1} = \vec{\theta}_t - \eta g_{\rm FS}^{-1}\frac{\partial{\cal L}}{\partial\vec{\theta}_t},
\end{equation}
where~$\eta$ is the learning rate~(typically a real number~$< 1$) and~$g_{\rm FS}$ is the Fubini-Study metric~\cite{Stokes2020, Roy2023efficient}. The latter is given by the real part of the quantum geometric tensor~$G$:
\begin{equation}
G = \bigg{\langle} \frac{\partial \psi_{\rm f}}{\partial\theta_i}\bigg{|}\frac{\partial \psi_{\rm f}}{\partial\theta_j}\bigg\rangle - \bigg\langle \frac{\partial \psi_{\rm f}}{\partial\theta_i}\bigg{|}\psi_{\rm f}\bigg\rangle\bigg\langle\psi_{\rm f}\bigg|\frac{\partial \psi_{\rm f}}{\partial\theta_j}\bigg\rangle.
\end{equation}
Note that both the gradients need to be evaluated using measurements for both the cost-function and the Fubini-Study metric~\cite{Wierichs_2022}.
Subsequently, the updated set of parameters are fed back to the quantum processor and the process is repeated until the iteration convergence criterion is satisfied. Once the latter is satisfied, a second layer convergence criterion is verified. This ensures that the number of layers chosen is sufficient to attain the desired degree of convergence. If this condition is not satisfied, the number of layers is grown by 1 and the earlier steps are repeated. Else, the algorithm terminates. When the number of layers is increased from $N$ to $N+ 1$, the optimized parameters obtained for the $N$-layer circuit, augmented by an initial guess for the added layer, were used as the choice for the optimization process for $N+1$ layers. This improved the stability of the algorithm.

For $N$ layers and a ring of~$L$ qubits, the total number of parameters determined by the above algorithm is~$10LN$, where the factor 10 originates from the number of free parameters in the two-qubit unitary operator of Fig.~\ref{fig:schematic}(e). The number of entangling CNOT gates per qubit is given by $3N$. Despite being tractable~[see Fig.~\ref{fig_3}(c) for demonstration], conservation of symmetries present in the target Hamiltonian can drastically reduce the number of parameters to be determined. Symmetries can be conserved by modifying the general circuit ansatz of Fig.~\ref{fig:schematic} and allowing only operators that respect this symmetry. Alternatively, the cost-function can be augmented to ensure that the obtained state lies in the correct symmetry sector. For example, states which are invariant under translation by~$r$ lattice site(s) can be obtained by starting with a state and circuit ansatz respecting this symmetry. For~$r = 1(2)$, the number of undetermined parameters for~$N$ layers reduces to~$10N~(20N)$. Alternatively, in order to obtain a $\pm1$ eigenstate of the translation operator~$T$, the cost-function can be augmented as:~${\cal L}(\vec{\theta})\rightarrow {\cal L}(\vec{\theta}) \mp \mu\langle \psi_{\rm f}(\vec{\theta})|T|\psi_{\rm f}(\vec{\theta})\rangle$, where~$\mu$ is a positive real number. Similar considerations apply to symmetry operators related to parity/particle number conservation. 

Note that the described circuit ansatz can be modified to incorporate constraints arising from specific hardware architectures. For instance, if all-to-all coupling is permitted (applicable, for instance, to trapped-ion based quantum computing platforms), the proposed ansatz can be straightforwardly generalized to take advantage of the additional couplings. In this work, the simplest case of only nearest-neighbor coupling is considered. 

\paragraph*{Results.} The proposed algorithm is benchmarked using three paradigmatic one-dimensional QFTs. The results were obtained by modifying standard time-evolution routines for matrix product states. In all three cases, the quantum processor considered comprises a periodic ring with a fixed number of qubits with nearest-neighbor connectivity. This is independent of the model under consideration. The results obtained using the optimized quantum circuit are compared with those from density matrix renormalization group~(DMRG) and exact computations. While the results shown below make use of the QNG optimizer, the latter is not essential for the proposed circuit ansatz. For instance, both Ref.~\cite{Rogerson2024} and Ref.~\cite{Roy:2024xdi} use ADAM~\cite{Kingma2017} or its variant for unitary manifolds for the current ansatz. 

First, the critical Ising model in a longitudinal magnetic field is analyzed. In the continuum limit, the QFT contains eight mesonic excitations which scatter with a diagonal scattering matrix~\cite{Zamolodchikov1989}. These mesons can be viewed as an extremal~($g\rightarrow 1$) limit of bound states formed by the Ising domain walls in the ferromagnetic phase~$g < 1$. The magnetic field leads to a linearly growing energy of separation of two domain walls~\cite{McCoy1978, Fonseca2006}, akin to what happens in quarks in quantum chromodynamics~\cite{tHooft1974}, see Fig.~\ref{fig_3}(c). The QFT describing this model is integrable and the number of mesons and their mass ratios are given by the rank and the exponents of the Lie algebra~$E_8$ respectively~\cite{Zamolodchikov1989}. The relevant lattice Hamiltonian, imposing periodic boundary condition, is:
\begin{equation}
\label{eq:H_I}
H_{\rm I} = -\sum_{j = 1}^L X_jX_{j+1} - g\sum_{j=1}^LZ_j - h\sum_{j=1}^LX_j,
\end{equation}
where~$g$ and~$h$ were chosen to be~$1$ and~$\simeq0.154$ respectively. Figs.~\ref{fig_2}(a) and \ref{fig_3}(a) together present the results for the lowest eight translation-invariant states obtained using the proposed algorithm. The circuit ansatz was chosen to be invariant under translation by 2 sites. Furthermore, the cost-function was augmented to ensure that only~$+1$ eigenstates of the operator for single-site translation were obtained. Fig.~\ref{fig_2}(a) shows the results for the ground state energy for system-sizes between $L = 8$ and 20 in steps of 4. The number of layers required to obtain a reasonable approximation of the ground state do not scale with the system-size. This is compatible with the observations of Ref.~\cite{Roy2023efficient} and is a consequence of the `relatively-low amount of entanglement' in ground states of one-dimensional models with gapped spectrum~\cite{Hastings2007}. In fact,~$N=2$ is sufficient to obtain  the ground state with error in the energy~$< 1\%$~[Fig.~\ref{fig_2}(d)]. Fig.~\ref{fig_3}(a) shows the results for the first seven excited states for~$L = 8$. For the chosen convergence criterion of~$10^{-3}$, the maximum error in the obtained energies was~$0.58\%$. In contrast to the ground state, the number of layers required to achieve this precision ranged between $L$ and $3L$. The different translation-invariant excited states can be classified as lattice counterparts of the single and multiple meson states existing in the QFT. The mass ratio for the lowest two mesonic excitations~[square and diamond markers in Fig.~\ref{fig_3}(a)] obtained for L~$=$ 8 obtained using the proposed algorithm is 1.39 which is close to the expected value of 1.41 obtained using exact diagonalization of  the corresponding spin-chain. This should be compared with the QFT prediction~\cite{Zamolodchikov1989} given by $2\cos(\pi/5)\simeq 1.618$. 

Next, the three-state Potts model in the presence of a magnetic field~\cite{Rutkevich:2009fr, Rutkevich:2014rha, Lencses:2015bpa, Rutkevich:2017qdw}  is analyzed. In contrast to its two-state counterpart described above, this model exhibits a richer state of excitations. The three states of each Potts spin are encoded in the triplet of states of two qubits: 
\begin{equation}
|0\rangle = |\uparrow\uparrow\rangle, |1\rangle = \frac{1}{\sqrt{2}}\left(|\uparrow\downarrow\rangle + |\downarrow\uparrow\rangle\right), |2\rangle = |\downarrow\downarrow\rangle.
\end{equation}
The relevant Hamiltonian is
\begin{equation}
\label{eq:H_P}
H_{\rm P} = -\frac{2}{3\sqrt{3}}\sum_{j  = 1}^{L'}\left(\sigma_j\sigma_{j+1}^\dagger + g\tau_j + h \sigma_j + {\rm h.c.}\right),
\end{equation}
where~$\sigma$ and~$\tau$ are the shift and clock matrices respectively~\cite{Mong:2014ova}. Periodic boundary condition was imposed and~$L' = L/2$ is the number of Potts spins encoded in~$L$ qubits. For~$g< 1, h = 0$, the model exhibits a ferromagnetic phase with three degenerate vacua and six domain walls~[the three shown in Fig.~\ref{fig_3}(d) and those obtained by reflection about the domain wall]. The presence of the magnetic field breaks the degeneracy of the ground space, giving rise to one true vacuum and two false vacua. The domain-walls experience a confining potential as in the Ising case. However, now there are not only mesonic excitations formed by two domain walls, but also baryonic excitations formed by three domain walls~[Fig.~\ref{fig_3}(d)]. Notice that~$H_{\rm P}$ commutes with the~$\mathbb{Z}_2$ charge operator~$C$ which acts trivially on the state~$|0\rangle$ and interchanges~$|1\rangle$ and~$|2\rangle$. Figs.~\ref{fig_2},\ref{fig_3} together show the results for~$g = h = 0.1$ for the lowest eight translation-invariant states with eigenvalue~$+1$ for the operator~$C$. The 2-site translation-invariant circuit ansatz was chosen with the~$C$-eigenvalue imposed using the cost-function. As in the Ising case, the ground state is reached with depth independent of the system-size, with errors in energy less than~$0.1\%$ for two layers~[Fig.~\ref{fig_2}(b, d)]. The excited states were obtained using between~$L$ and~$3L$ layers with maximum error in the obtained energies being~$2.26\%$~[Fig.~\ref{fig_3}(b)]. The first excited state is one of the false vacua, with the rest comprising mesonic and baryonic excitations. This is verified by counting the average number of domain-walls in the corresponding state. 

Finally, results are presented for the canonical model for quantum electrodynamics in two space-time dimensions: the one-flavor massive Schwinger model~\cite{Coleman:1975pw, Banks1976}. After mapping to an array of qubits, the resulting Hamiltonian with open boundary conditions is given by~\cite{Banuls:2013jaa, Farrell2024}:
\begin{align}
H_{\rm MS} &= \frac{1}{2}\sum_{j = 1}^{L  - 1}\left(X_jX_{j+1} + Y_jY_{j+1}\right) + m\sum_{j = 1}^LP_j \nonumber\\\label{eq:H_MS}&\quad+ \frac{g^2}{2}\sum_{j = 1}^{L - 1}\left[\sum_{k =1}^j (-1)^kP_k\right]^2,
\end{align}
where~$P_j = [1 + (-1)^jZ_j]/2$. In contrast to the Ising and Potts models, this model involves an all-to-all coupling Hamiltonian. To obtain the ground state of this model, the general circuit ansatz of Fig.~\ref{fig:schematic} is used. Note the existence of the conserved U(1) charge~$Q$,~$[H, Q] = 0$, where~$Q = \sum_{j=1}^{L}Z_j$. To test the algorithm under the most challenging of conditions, the ground states were obtained for system sizes between L = 8 and 16 without conserving this charge. For these simulations,~$m = 0.5, g = 0.3$ were chosen. Despite the long-range nature of the model Hamiltonian, reasonable approximations of the ground-state were obtained with circuit-depths that are independent of the system-size with energies within~$1\%$ accuracy obtained within application of 2 layers. Excited states were also obtained similar to the Ising and Potts models and are not shown here for brevity. In the three examples analyzed above, the number of iterations required for convergence for a given number of layers is~${\cal O}(10^2)$. 

\paragraph*{Conclusion.} To summarize, a hybrid quantum-classical algorithm is presented for the investigation of strongly-interacting QFTs. In contrast to existing approaches that rely on heuristic arguments for choosing the parametrized quantum circuit ansatz, a universal ansatz is described relying on Euler's and Cartan's decomposition of single and two-qubit operators. The algorithm is benchmarked by computing eigenstates of lattice models for three paradigmatic one-dimensional QFTs: the Ising, Potts and one-flavor massive Schwinger models. Quantum circuits were obtained for the realization of ground states, false vacua, mesonic and baryonic excitations occurring in these models. 

The presented results demonstrate the power of geometric control theory~\cite{Jurdjevic1997} for the analysis of QFTs. While geometric quantum control is known to be the central concept in the search for an optimal circuit for quantum computation~\cite{Nielsen2005, Nielsen2006, Dowling2006}, previous proposal are centered around minimization of the distance in the Finsler space of unitary operators. In contrast, the current work presents an alternative framework based on the Fubini-Study metric and a universal circuit ansatz. The proposed framework is not only numerically-tractable, but also readily-implementable on current quantum simulators, enabling efficient preparation of multiple-meson or domain-wall excitations of confining QFT models in two space-time dimensions. These can shed novel insights into string fragmentation dynamics in models emulating quantum chromodynamics as well as scattering amplitudes in non-integrable perturbations of two-dimensional conformal field theories. 

\paragraph*{Acknowledgements.}
This work was supported by the U.S. Department of Energy, Office of Basic Energy Sciences, under Contract No. DE-SC0012704. 

\bibliography{/Users/ananda/Dropbox/Bibliography/library_1}

\end{document}